\documentclass[pre,twocolumn,showpacs]{revtex4-1}
\usepackage[dvips]{graphicx,psfrag}
\usepackage{amsfonts}
\usepackage{setspace}
\usepackage{mathrsfs}
\usepackage[dvips,usenames]{color}
\usepackage{amsmath,amssymb,enumerate}
\usepackage{ulem}

\begin{document}


\title[First passage times in bounded domains]{First passages in bounded
domains: When is the mean first passage time meaningful?}

\author{Thiago G. Mattos}
\email{Corresponding author: tgmattos@is.mpg.de}
\affiliation{Max Planck Institute for Intelligent Systems, Heisenbergstr. 3,
70569 Stuttgart, Germany.}
\author{Carlos Mej\'{\i}a-Monasterio}
\affiliation{Laboratory of Physical Properties,
Technical University of Madrid, Av. Complutense s/n, 28040 Madrid, Spain.}
\affiliation{Department of Mathematics and Statistics,
University of Helsinki,\\P.O. Box 68, FIN-00014 Helsinki, Finland}
\author{Ralf Metzler}
\affiliation{Institute for Physics \& Astronomy, University of Potsdam, 14476
Potsdam-Golm, Germany \& Physics Department,  Tampere University of
Technology, Korkeakoulunkatu 3, FI-33101 Tampere, Finland.}
\author{Gleb Oshanin}
\affiliation{Laboratoire de Physique Th{\'e}orique de la Mati{\`e}re
Condens{\'e}e (UMR CNRS 7600), Universit{\'e} Pierre et Marie Curie (Paris 6) -
4 Place Jussieu, 75252 Paris, France.}


\begin{abstract}
We study the first passage statistics to adsorbing boundaries of a Brownian
motion in bounded two-dimensional domains of different shapes and
configurations of the adsorbing and reflecting boundaries. From extensive
numerical analysis we obtain the probability $P(\omega)$ distribution of the
random variable $\omega=\tau_1/(\tau_1+\tau_2)$, which is a measure for how
similar the first passage times $\tau_1$ and $\tau_2$ are of two independent
realisations of a Brownian walk starting at the same location. We construct
a chart for each domain, determining whether $P(\omega)$ represents a unimodal,
bell-shaped form, or a bimodal, M-shaped behavior. While in the former case
the mean first passage time (MFPT) is a valid characteristic of the first
passage behavior, in the latter case it is an insufficient measure for
the process. Strikingly we find a distinct turnover between the two modes of
$P(\omega)$, characteristic for the domain shape and the respective location
of absorbing and reflective boundaries. Our results demonstrate that large
fluctuations of the first passage times may occur frequently in two-dimensional
domains, rendering quite vague the general use of the MFPT
 as a robust measure of the actual behavior
 even in bounded domains, in which all moments of the first passage
distribution exist.
\end{abstract}

\date{\today}

\maketitle


\section{Introduction}
\label{sec:intro}

The concept of first passage underlies diverse stochastic processes in which
it is relevant when the value of the random variable reaches a preset value
for the first time. A few stray examples across disciplines include chemical
reactions \cite{ol,lov,ben,beni,mattosreis,smolu}, the firing of a neuron
\cite{gersh,burkitt}, random search of a mobile or immobile target
\cite{alb,alb1,ol1,ralf,osh,wio,carlos,paul,port,evans,gel}, diffusional disease
spreading \cite{Lloyd2001}, DNA bubble breathing \cite{hanke}, dynamics
of molecular motors \cite{motor1,motor2}, the triggering of a stock
option \cite{bouch}, etc. A variety of first passage time phenomena and
different related results have been investigated in Refs.~\cite{katja,sid}.
While for continuous processes the first passage across a given preset
value coincides with the first arrival to exactly this value, for L{\'e}vy
flights characterized by long-tailed jump length distributions with
diverging variance both quantities become different, and large overshoots
across a preset value occur \cite{chech}.

The distribution of first passage times in unbounded domains is typically
broad, such that not even the mean first passage time exists \cite{sid}.
In particular, in one-dimensional, semi-infinite domains the first passage
time distribution of a Markovian process is universally dominated by the
$t^{-3/2}$ scaling nailed down by the Sparre Andersen theorem \cite{sid}.
A similar divergence of the mean first passage time occurs in stochastic
processes characterized by scale-free distributions of waiting times
\cite{mekla}. In contrast, in many practically important situations first
passage processes involve particles which move randomly in bounded domains
(see, e.g., Refs.~\cite{ol2,ol3,ol4,fBM-wedge}). In this case the random
variable of interest, the first passage time $\tau$ to, e.g., a boundary,
a target chemical group, a binding site on the surface of the domain or
elsewhere within the domain, etc., has  a distribution $\Psi(\tau)$ of the
generic, generalized inverse Gaussian form (see {\it e.g.}, the discussion in Ref.~\cite{carlos} and references therein)
\begin{equation}
\label{eq:Pt-exptrunc}
\Psi(\tau)\sim \exp\left(-\frac{a}{\tau}\right)\frac{1}{\tau^{1+\mu}}\exp
\left(-\frac{\tau}{b}\right),
\end{equation}
where $a$ and $b$ are some constants dependent on the shape of the domain,
the exact starting point within the domain, etc., and $\mu$ is the so-called
persistence exponent \cite{satya}. When the linear size of the domain (say, the radius $R$ of a circular or a spherical domain)
diverges (i.e., $R\to\infty$), the parameter $b$ also diverges such that the
long-time asymptotic behavior of the first passage time distribution is of
power-law form without a cutoff. In this case, at least some, if not all, of
the moments of $\Psi(\tau)$ diverge.

The first passage time distribution in Eq.~(\ref{eq:Pt-exptrunc}) is
exact only in the particular case of Brownian motion on a
semi-infinite line in presence of a bias pointing towards the target
site, or, equivalently, for the celebrated integrate-and-fire model of
neuron firing by Gerstein and Mandelbrot \cite{gersh}. In general, the
detailed form of $\Psi(\tau)$ is obviously much more complex than
given by Eq.~(\ref{eq:Pt-exptrunc}), depending on the very shape of
the domain under consideration and the exact boundary value
problem. Typically $\Psi(\tau)$ is given in terms of an infinite
series. Nonetheless, on a \textit{qualitative\/} level, the
approximation (\ref{eq:Pt-exptrunc}) provides a clear picture of the
actual behavior of the first passage time distribution in bounded
domains. Namely, $\Psi(\tau)$ consists of three different parts: a
singular decay for small values of $\tau$, which mirrors the fact that
the first passage to some point starting from a distant position
cannot occur instantaneously. This is followed at intermediate times
by a generic power-law decay with exponent $\mu$, depending on the
exact type of random motion.  Finally, an exponential decay at long
$\tau$ cuts off the power-law. A crucial aspect is that the
exponential cutoffs at both short and long $\tau$ ensure that in
bounded domains $\Psi(\tau)$ possesses moments of arbitrary positive
or negative order.

Distributions of the form (\ref{eq:Pt-exptrunc}) are usually considered
\textit{narrow}, as opposed to \textit{broad\/} distributions, which do not
possess all moments \cite{katja,sid,mekla,chech}, e.g., $\Psi(\tau)$ in
Eq.~(\ref{eq:Pt-exptrunc}) with $b=\infty$. Once all moments exist, it is
often tacitly assumed that the first moment of this distribution, the mean
first passage time (MFPT)
\begin{equation}
\label{mfpt}
\langle\tau\rangle=\int_0^{\infty}\tau\Psi(\tau)d\tau,
\end{equation}
is an adequate measure of the first passage behavior. The actual
analytical calculation of the MFPT may require a considerable
computational effort, and the calculation of higher moments is quite
formidable and is not always possible, (compare, e.g.,
Refs.~\cite{ol2,ol3,ol4}). Conversely, it has been demonstrated in,
e.g., recent Refs.~\cite{red,samor,samor2,hol,schehr,gl} that random variables
with truncated power-law distributions behave in several important
aspects as those characterized by non-truncated, \textit{broad\/}
distributions, revealing substantial fluctuations between individual
realisations and thus rendering the concept of a mean first passage
time a bit unsubstantiated.  To be more precise, this concerns not the
functional form of the MFPT for a given process, but rather its use as
a characteristic quantity for the process.  The functional form of the
MFPT is certainly an important property, providing valuable insights
to the scaling behavior, for instance with the system size or the
initial distance of starting point and target. In contrast, the very
numerical value of the MFPT can significantly differ from the values
drawn from individual trajectories. Therefore, the MFPT can be
substantially larger than the most probable value for the first
passage time.  Clearly, an understanding of how representative the
MFPT is of the actual behavior and, concurrently, how important
fluctuations of $\tau$ between individual realisations indeed are of
utmost conceptual importance in many areas, such as, e.g., an
interpretation of the first passage data obtained from single particle
tracking.

In this paper we analyze, via extensive Monte Carlo simulations the role of
fluctuations between individual realisations of first passage times for
Brownian motion (BM) in two-dimensional bounded domains of different shapes,
and with different configurations of the reflective and adsorbing boundaries.
Analogous results for three-dimensional systems and for systems with quenched
disorder will be presented elsewhere \cite{thiago2}.

\section{Simultaneity concept of first passage}

\begin{figure}
\centerline{\includegraphics*[width=0.75\linewidth]{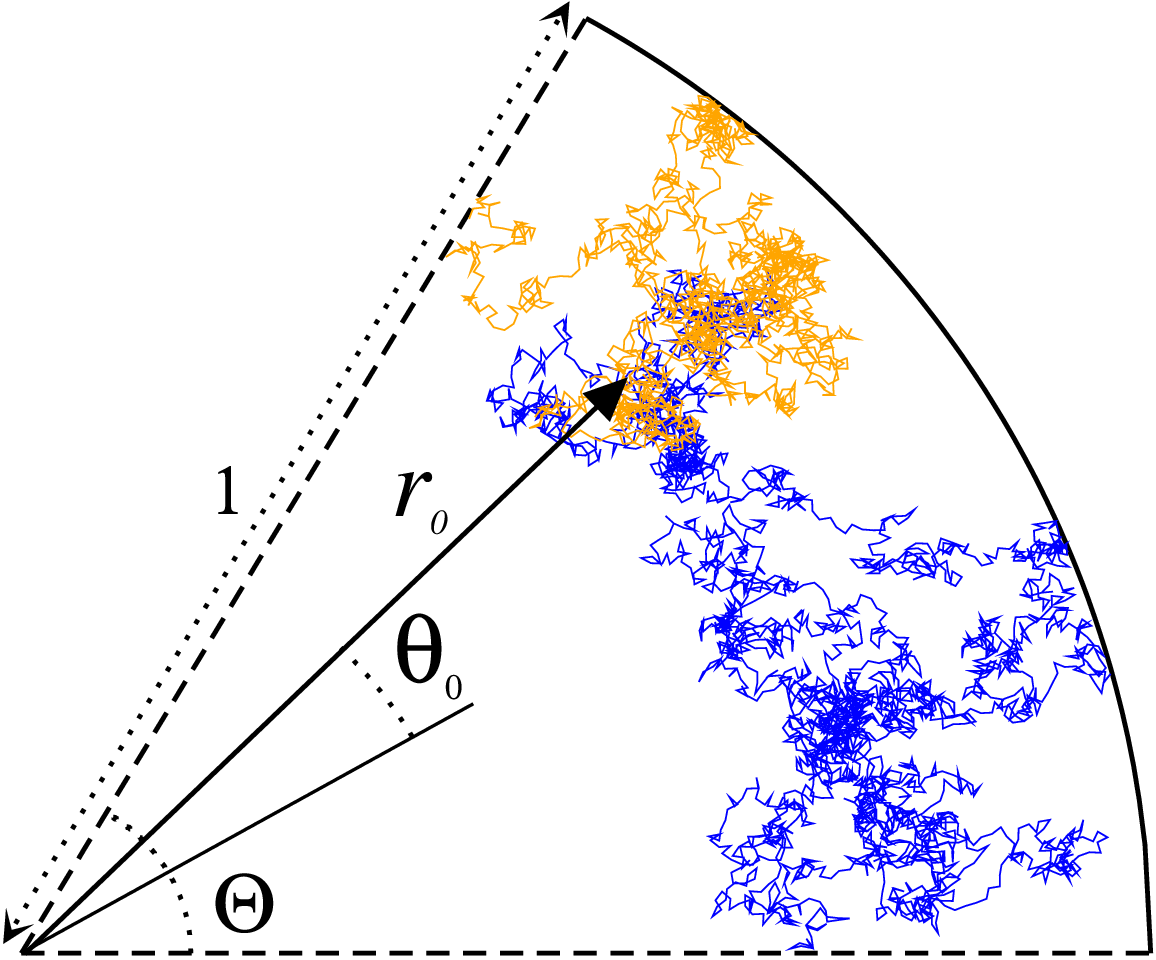}}
\caption{(Color online) Trajectories of two Brownian walkers starting at the same initial
position $(r_0,\theta_0)$ inside a bounded pie-wedge domain with opening angle
$\Theta$ as well as absorbing radial boundaries (dashed lines) and reflecting
boundary (solid line) at $r=1$. The values of the first passage times
to the adsorbing boundaries are used to construct the random variable $\omega$.}
\label{fig:scheme}
\end{figure}

To quantify the relevance of such fluctuations and the effective \textit{broadness\/} of
the corresponding first passage time distribution $\Psi(\tau)$ we employ a
novel diagnostics method based on the concept of simultaneity of first passage
events, compare Fig.~\ref{fig:scheme}. Instead of the original first passage
problem with quantifying the statistical outcome for a single Brownian walker,
we simultaneously launch two identical, non-interacting Brownian particles
at the same position ${\bf r_0}$ (which is identical to two different
realisations of the trajectories of a single BM starting at ${\bf r_0}$). The
corresponding outcomes are the first passage times $\tau_1$ and $\tau_2$. We
now define the random variable
\begin{equation}
\label{def:omega}
\omega\equiv\frac{\tau_1}{\tau_1+\tau_2},
\end{equation}
such that $\omega$ ranges in the interval $[0,1]$. The \textit{uniformity
index\/} $\omega$ measures the \textit{likelihood\/} that both walkers arrive to
the adsorbing boundary simultaneously: when $\omega$ is close to 1/2, the
process is uniform and the particles behave as if they were
almost performing a Prussian \textit{Gleichschritt}. In contrast,
values of $\omega$ close to 0 or 1 mean highly non-uniform behavior, implying
that the MFPT is not representative of the actual behavior,
but is merely the first moment of an \textit{effectively} broad
distribution. We note parenthetically that similar random variables have been used in the analysis of random probabilities induced by normalization of self-similar L\`evy processes~\cite{iddo1}, of the fractal characterization of Paretian Poisson processes~\cite{iddo2}, and of the so-called Matchmaking paradox~\cite{iddo3,iddo4}.

Within a given bounded domain we evaluate the distribution $P(\omega)$
measuring the uniformity of the first passage dynamics with respect to
some fixed starting point $\bf r_0$. This is repeated for a large
number of nodes $\bf r_0$ within the domain, thus producing a
uniformity chart of first passage. Remarkably, we find that the very
shape of this distribution depends delicately on the domain shape, the
actual settings of adsorbing and reflecting boundaries, and on the
starting location $\bf r_0$. In some starting areas $P(\omega)$ has a
characteristic unimodal, bell-shaped form with a maximum at $\omega
=1/2$, signaling that most pairs of BMs will arrive to the adsorbing
boundary simultaneously.  This means, in turn, that in this case the
parental first passage time distribution $\Psi(\tau)$ can be
considered as sufficiently \textit{narrow\/} such that the MFPT can be
considered as a plausible measure of individual first passage events,
providing a rather accurate estimate for the typical value of the
first passage time. Conversely, we find that for other starting areas
$P(\omega)$ exhibits a completely different behavior and has a
characteristic bimodal, M-shaped form with a local \textit{minimum\/} at
$\omega=1/2$ and two maxima close to $\omega=0$ and $\omega=1$. In that case simultaneous arrival of two initially
synchronized walkers is unlikely, i.e., any two trajectories will most
likely possess distinctly different first passage times. The parental
first passage time distribution $\Psi(\tau)$ is consequently
\textit{broad\/} and sample-to-sample fluctuations matter: the MFPT
cannot be considered as an adequate measure of the actual behavior.
Given that, by definition, the averages $\langle\tau_1\rangle$ and
$\langle \tau_2\rangle$ are identical and, moreover, the moments of
$\tau_1$ and $\tau_2$ of arbitrary order coincide, one can think of
$\omega$ (and, hence, of the distribution $P(\omega)$) as a measure of
the symmetry breaking between different realisations of the
process. Note also that situations in which the mean value of some
pertinent parameter is dominated by the tails of the distribution, and
this mean thus has a very different value compared to the most
probable value (and may even show a completely different dependence on
the system parameters) is most often encountered in disordered systems
\cite{evans,gl}.  Here we observe such a behavior in absence of any
disorder.

Scanning then over the
possible starting points within each bounded domain, we obtain a corresponding
phase-chart for $P(\omega)$, distinguishing regions in which $P(\omega)$ has
M-shaped or bell-shaped behavior. The demarcation zone between these two
phases, depicted by beige color in the following, represents a plateau-like,
almost uniform behavior of $P(\omega)$ with zero second derivative at
$\omega=1/2$.

We proceed by giving a general definition of the first passage time
distribution $\Psi(\tau)$ and its corresponding MFPT in Section \ref{sec:MFPT},
and also establish a relation between $\Psi(\tau)$ and the uniformity
distribution $P(\omega)$. In Section \ref{sec:pie} we study in detail the
problem of Brownian motion in a pie-wedge shaped domain with absorbing and
reflecting boundaries. In Sections~\ref{sec:circle} and \ref{sec:triangle}
we discuss the forms of $P(\omega)$, as a function of the location of the
starting point, for circular domains with small aperture on the boundary,
a two-dimensional version of the so-called Narrow Escape Time problem \cite{NET},
and for triangular domains with adsorbing boundaries, respectively. Our
results are summarized in Section \ref{sec:concl}.

\section{First passage distribution, mean first passage time, and the
uniformity distribution $P(\omega)$.}
\label{sec:MFPT}

Consider a BM inside a general two-dimensional domain $\mathcal{S}$, whose
boundary $\partial\mathcal{S}\equiv\partial\mathcal{S}_a\cup\partial\mathcal{
S}_r$ comprises reflecting, $\partial\mathcal{S}_r$, and absorbing, $\partial
\mathcal{S}_a$, parts. At time $t=0$, the BM initiates at $\mathbf{r}_0\in
\mathcal{S}$ and evolves within the domain until the trajectory hits $\partial
\mathcal{S}_a$ for the first time at some random instant $\tau$. Furthermore let
$P(\mathbf{r},t|\mathbf{r}_0)$ denote the conditional probability distribution
for finding the Brownian walker at position $\mathbf{r}$ at time $t$, provided
the initial condition was at $\mathbf{r}_0$ at $t=0$. The distribution $P(
\mathbf{r},t|\mathbf{r}_0)$ is the solution of the diffusion equation
\begin{equation}
\label{diff}
\frac{\partial}{\partial t}P(\mathbf{r},t|\mathbf{r}_0)=D\nabla^2_{\mathbf{r}}
P(\mathbf{r},t|\mathbf{r}_0)
\end{equation}
on $\mathcal{S}$, where $\nabla^2_{\mathbf{r}}$ is the two-dimensional
Laplacian equivalent to $\partial^2/\partial x^2+\partial^2/\partial y^2$ in
Cartesian coordinates. Eq.~(\ref{diff}) is subject to the initial condition
as well as the boundary conditions at $\partial\mathcal{S}$. Here $D$ is the
diffusion coefficient. The solution of this boundary value problem is, in the
best case, cumbersome, and explicit solutions may be obtained for only few
simple geometries, compare Ref.~\cite{Carslaw}.

If a finite part of the boundary is absorbing, i.e.,
$\partial\mathcal{S}_a$ is not empty, then the distribution
$P(\mathbf{r},t|\mathbf{r}_0)$ is no longer normalized.  The survival
probability $\mathscr{S}_{\mathbf{r}_0}(t)$ that the walker has not
reached $\partial\mathcal{S}_a$ up to time $t$, is defined by
\begin{equation}
\label{surv}
\mathscr{S}_{\mathbf{r}_0}(t)=\int_{\mathcal{S}}P(\mathbf{r},t|\mathbf{r}_0)
d\mathbf{r}.
\end{equation}
$\mathscr{S}_{\mathbf{r}_0}(t)$ is a monotonically decreasing function
of time, eventually reaching zero value,
$\lim_{t\to\infty}\mathscr{S}_{\mathbf{r} _0}(t)=0$. The desired
distribution of first passage times to the adsorbing boundary becomes
\begin{equation}
\label{FPT}
\Psi_{\mathbf{r}_0}(\tau)=-\frac{d\mathscr{S}_{\mathbf{r}_0}(\tau)}{d\tau}.
\end{equation}
The MFPT associated with the distribution $\Psi(\tau)$ is defined as the first
moment
\begin{equation}
\label{MFPT}
\langle\tau\rangle(\mathbf{r}_0)=\int_0^\infty\tau\Psi_{\mathbf{r}_0}(\tau)
d\tau=\int_0^\infty\mathscr{S}_{\mathbf{r}_0}(\tau)d\tau.
\end{equation}
We note parenthetically that in most of the existing literature,
apart of recent Refs.\cite{ben,ol2,ol3},
the dependence
of the MFPT on the starting position of the walker is either simply neglected,
or it is assumed that the starting point is randomly distributed within the
domain $\mathcal{S}$. As we proceed to show, the $\mathbf{r}_0$-dependence
of the first passage time distribution is a crucial aspect which cannot be
neglected.

We now turn to the uniformity distribution $P(\omega)$ of the random variable
$\omega$, Eq.~(\ref{def:omega}). Let
\begin{equation}
\label{mg}
\Phi(\lambda)=\int_0^1P(\omega)\exp\left(-\lambda\omega\right)d\omega,
\end{equation}
with $\lambda\geq0$, denote the moment generating function of $\omega$. Since
$\tau_1$ and $\tau_2$ are independent, identically distributed random variables,
expression (\ref{mg}) can formally be represented as
\begin{equation}
\label{2}
\Phi(\lambda)=\int^{\infty}_0\!\!\!\!\int^{\infty}_0\!\!\!\!\Psi(\tau_1)\Psi(\tau_2)\exp\left(
-\lambda\frac{\tau_1}{\tau_1+\tau_2}\right)d\tau_1d\tau_2.
\end{equation}
Integrating over $d\tau_1$ we change the integration variable, $\tau_1\to
\omega$, so that Eq.~(\ref{2}) is rewritten in the form
\begin{eqnarray}
\label{3}
\Phi(\lambda) & = & \int^{1}_0\exp\left(-\lambda\omega\right)\frac{d\omega}{(1-
\omega)^2}\times\nonumber \\
 & & \int^{\infty}_0\tau_2\Psi(\tau_2)\Psi\left(\frac{\omega}{1-\omega}\tau_2\right)
d\tau_2.
\end{eqnarray}
From comparison with Eq.~(\ref{mg}), we readily read off the desired
distribution function
\begin{equation}
\label{def:Pw}
P(\omega)=\frac{1}{(1-\omega)^2}\int^{\infty}_0\tau\Psi(\tau)
\Psi\left(\frac{\omega}{1-\omega}\tau\right)d\tau.
\end{equation}
Therefore, $P(\omega)$ is known for given $\Psi(t)$.

To get an idea of the typical behavior of the uniformity distribution $P(
\omega)$, we use the generic form (\ref{eq:Pt-exptrunc}) for the first passage
time distribution. From Eq.~(\ref{def:Pw}) we find from integration that
\begin{eqnarray}
\label{eq:Pw-exptrunc}
P(\omega) & = & \frac{1}{2K_{\mu}^2\left(2\sqrt{\frac{a}{b}}\right)}
\frac{1}{\omega(1-\omega)}\times \nonumber \\
 & & K_{2\mu}\left(2\sqrt{\frac{a}{b\omega
(1-\omega)}}\right),
\end{eqnarray}
where $K_{2 \mu}(\cdot)$ is the modified Bessel function of the second type.
It was realized \cite{carlos,schehr} that the form of the distribution
$P(\omega)$ in Eq.~(\ref{eq:Pw-exptrunc}) is distinctly sensitive to the
value of the persistence exponent $\mu$, which characterizes the scaling
behavior of the first passage time distribution $\Psi(\tau)$ at intermediate
times. Thus, for $\mu>1$, $P(\omega)$ is always a unimodal, bell-shaped
function with a maximum at $\omega=1/2$. For $\mu=1$, $P(\omega)$ is almost
uniform, $P(\omega)\approx1$, apart from narrow regions at the corners $\omega
=0$ and $\omega=1$, for $b/a\gg1$. Curiously, for $\mu<1$, which corresponds to
the most common case, there exists a critical value $p_c$ of the ratio $p=b/a$
such that for $p>p_c$ the distribution $P(\omega)$ has a characteristic
M-shaped form with two maxima close to 0 and 1, while at $\omega=1/2$ we
find a local minimum. Such a transition from a unimodal, bell-shaped to bimodal,
M-shaped form mirrors a significant manifestation of sample-to-sample
fluctuations that has been indeed observed in exact calculations of $P(\omega)$
for Brownian search processes for an immobile target in $d$-dimensional
spherical geometries \cite{carlos}.

In what follows we further explore this intriguing behavior of the first
passage time distribution via extensive Monte Carlo simulations focusing
on the effects of the domain shape, the type of the boundary conditions,
and the initial position of the walker.

\section{Uniformity distribution $P(\omega)$ in a pie-wedge domain}
\label{sec:pie}

Consider now the case of a bounded domain of pie-wedge shape with unit radius,
$R=1$ and opening angle $\Theta$. The absorbing boundaries correspond to the
radial edges, while the outer circular edge is reflective, compare
Fig.~\ref{fig:scheme}.  Clearly, for a BM inside such a pie-wedge domain, all
moments of the first passage time distribution exist.

Before we proceed to investigate this case we first turn to the case
when the wedge radius is infinite, $R\rightarrow\infty$. Then the
distribution function $P(\mathbf{r},t|\mathbf{r}_0)$ is known exactly,
(see, e.g., Refs.~\cite{sid,metzler}) and is represented by an infinite
series whose leading term for $t\to\infty$ is given, up to a
normalization constant, by
\begin{eqnarray}
\label{eq:Pinf}
P(\mathbf{r},t|\mathbf{r}_0) & \simeq & \frac{\pi\sin\left(\pi\theta_0/\Theta\right)}{
4D\Theta t}e^{-(\rho^2+\rho_0^2)/4Dt}\times \nonumber \\
 & & I_{\pi/\Theta}\left(\frac{\rho_0 \rho}{2Dt} \right),
\end{eqnarray}
where $I_\nu(z)$ is the modified Bessel function of the first kind and ${\bf r}=(\rho,\theta)$ is
conveniently represented in polar coordinates. This solution is obtained for the
sharp initial condition $P(\mathbf{r},0|\mathbf{r}_0)=\pi\sin(\pi\theta_0/\Theta)
\delta({\bf r}-{\bf r_0})/2\Theta\rho_0$. From Eq.~(\ref{eq:Pinf}) one finds the
asymptotic behavior of the survival probability,
\begin{equation}
\label{eq:Sinf}
\mathscr{S}_{\mathbf{r}_0}(t)\simeq\left(\frac{\rho_0^2}{D}\right)^{\pi/2\Theta}
t^{-\pi/2\Theta},
\end{equation}
such that the first passage time distribution becomes
\begin{equation}
\label{eq:FPTDinf}
\Psi_{\mathbf{r}_0}(t)\simeq\frac{\pi}{2\Theta}\left(\frac{\rho_0^2}{D}\right)^{
\pi/2\Theta}\frac{1}{t^{1+\pi/2\Theta}}.
\end{equation}
Note that this distribution is of the generic form (\ref{eq:Pt-exptrunc}),
where $b=\infty$ due to the infinite domain size. The non-universal persistence
exponent is given by
\begin{equation}
\mu=\frac{\pi}{2\Theta}.
\end{equation}
Therefore, the MFPT diverges when $\Theta\geq \pi/2$ and is finite for $\Theta<\pi
/2$. According to the qualitative analysis from Section \ref{sec:MFPT}, $P(
\omega)$ will have a bimodal form in the former case and a unimodal one in the
latter.

We now turn our attention to finite-sized pie-wedges, for which the MFPT and all
higher moments of Eq.~(\ref{FPT}) are finite. In principle, an exact solution
for the first passage time distribution in this case can be obtained by
solution of the corresponding mixed boundary value problem, but the result will
be too cumbersome for our purposes. Instead, we resort to numerical
simulations. We now show that for finite pie-wedges the actual behavior
is in fact richer than in the case of an infinite wedge.

We performed Monte Carlo simulations of a random walk inside a pie-wedge of
unit radius and opening angle $\Theta$. The boundary conditions along the
radii are absorbing and reflecting along the circular edge,
see Fig.~\ref{fig:scheme}. The random walk is simulated in terms of a standard
Pearson walk on a plane (compare Ref.~\cite{Hughes}), which consists of a
sequence of steps of fixed length $\lambda=0.001$ and uniform waiting time
$v=1/\lambda$. After each step the walker turns by a random angle with uniform
distribution. At time $t=0$ the walker is released at $(\rho_0,\theta_0)$, and
its trajectory is recorded until it hits a point on the absorbing boundary for
the first time. Generating $N$ (we used $N=10^5$) such trajectories, we obtain
a set of first passage times $\{\tau_i\}$, from which we construct the first
passage time distribution. Since all $\tau_i$ are  independent, identically
distributed random variables, the uniformity distribution $P(\omega)$ is then
readily obtained via Eq.~\eqref{def:omega} from distinct pairs $\tau_1$ and
$\tau_2$ chosen at random from the set $\{\tau_i\}$.

\begin{figure}
\centerline{\includegraphics*[width=0.5\linewidth]{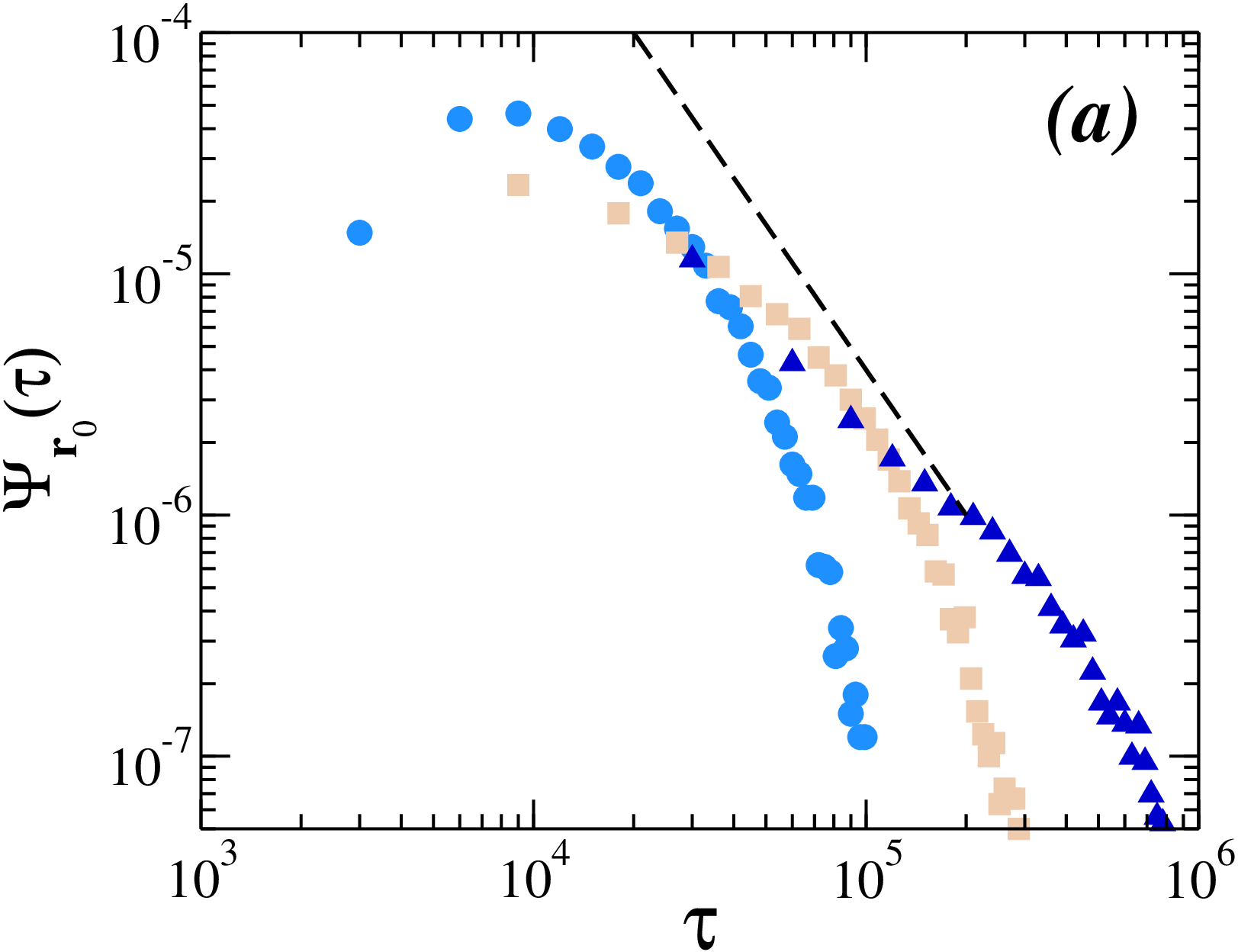}\includegraphics*[width=0.5\linewidth]{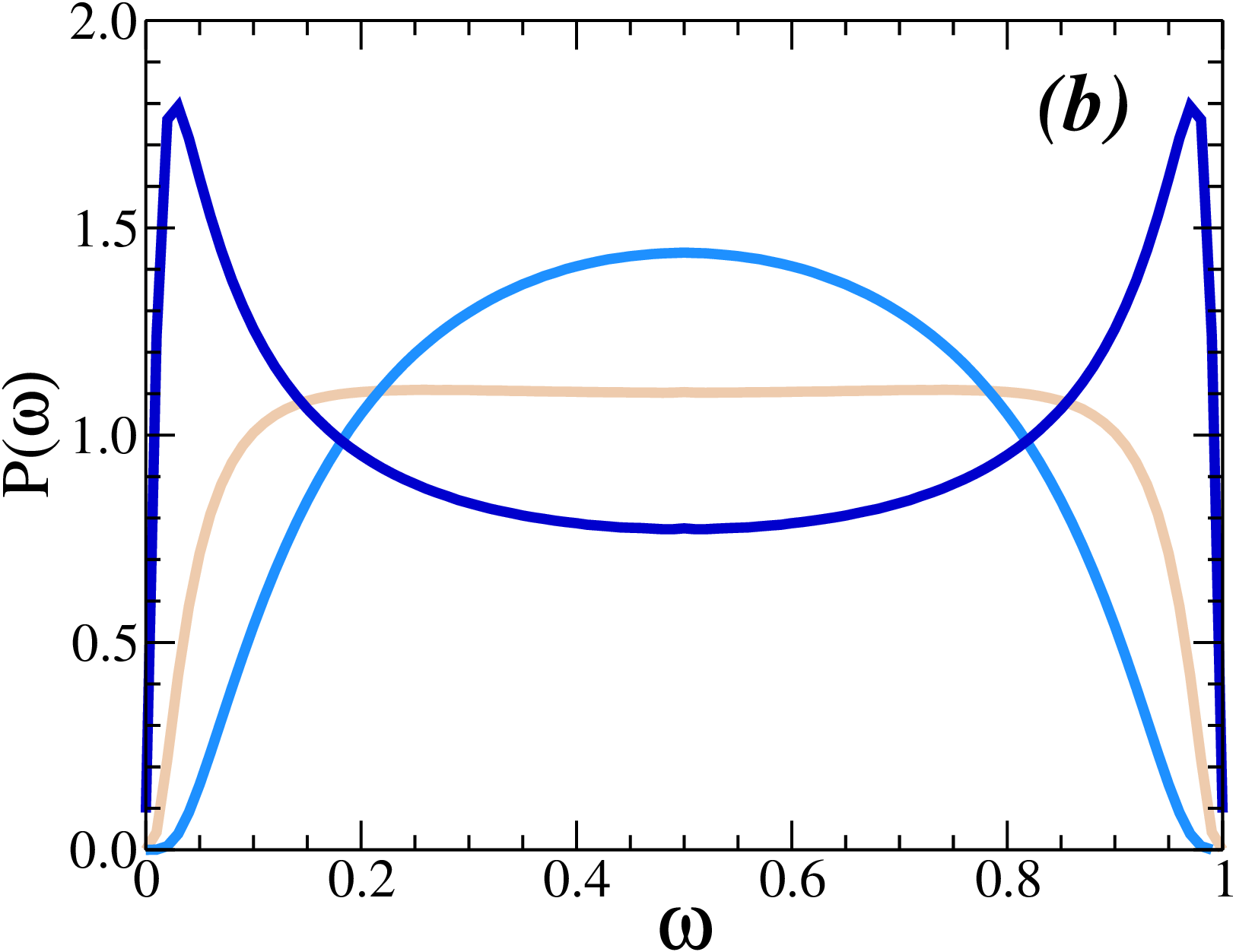}}
\caption{(Color online)(a) First passage time distribution $\Psi_{(\rho_0,\theta_0)}(\tau)$
for a Brownian walk in a pie-wedge domain with opening angle $\Theta=\pi/2$.
Different colors and symbols correspond to different starting points: $(\rho_0=0.76,
\theta_0=-0.38~\Theta$ (dark blue/triangles), $(0.76,-0.23~\Theta)$ (beige/squares), and $(0.76,0)$
(light blue/circles). The dashed straight line indicates the intermediate power-law
decay $\Psi(\tau)\sim1/\tau^2$, Eq.~(\ref{eq:FPTDinf}). (b). The
corresponding distribution $P(\omega)$ with the same color and symbol coding. In grey scale, dark blue corresponds to the darkest shade of grey, light blue to dark grey and beige to light grey.}
\label{fig:pie-fptd}
\end{figure}

Fig.~\ref{fig:pie-fptd}(a) shows the first passage time distributions
corresponding to a fixed $\rho_0=0.76$ and three different starting
angles $\theta_0$ for a pie-wedge  with opening angle $\Theta=\pi/2$.
One notices that, for small and large values of $\tau$, all three distributions
$\Psi(\tau)$ significantly deviate from the intermediate power-law behavior, which is 
due to exponential tempering. On the other hand, at intermediate times the distributions
exhibit a slower, power-law like decay within a range that depends
significantly on $\theta_0$. The narrowest distribution (light blue)
is obtained for a starting position $(0.76,0)$, which is exactly on the
symmetry axis of the wedge. Increasing the  angle $\theta_0$ away from the
symmetry axis results in a broadening of $\Psi(\tau)$, and the intermediate
algebraic decay is more pronounced.

In panel (b) of Fig.~\ref{fig:pie-fptd} we plot the corresponding uniformity
distributions $P(\omega)$. For the narrowest first passage time distribution
$\Psi(\tau)$ (light blue symbols), $P(\omega)$ is  bell-shaped with 1/2 representing
the most probable value, such that sample-to-sample fluctuations of $\tau$ are
less significant. In this case, apparently, the MFPT is a meaningful, 
reliable measure of
the first passage behavior, and any two walkers starting from the same position
on the symmetry axis of the wedge will most likely be absorbed at the same
instant of time. Strikingly, we find that this is no longer valid for the two
other starting positions off the symmetry axis: for ${\bf r_0}=(0.76,-\pi/3)$
the uniformity distribution $P(\omega)$ is almost uniform, except for narrow
regions in the vicinity of the edges, meaning that any relation between
the first passage times of two walkers is equally probable. Finally, for
starting position $(0.76,-\pi/7)$, which is the one closest to the absorbing
boundary $P(\omega)$ has a characteristic M-shaped form with maxima close to
0 and 1, and a local minimum at $\omega=1/2$. This signifies that in a such a
case the symmetry between any two walkers is most distinctly broken, and they
will be absorbed at very different times. Clearly, in this case the MFPT is
not representative of the actual behavior.

\begin{figure}
\centerline{\includegraphics*[width=1.0\linewidth]{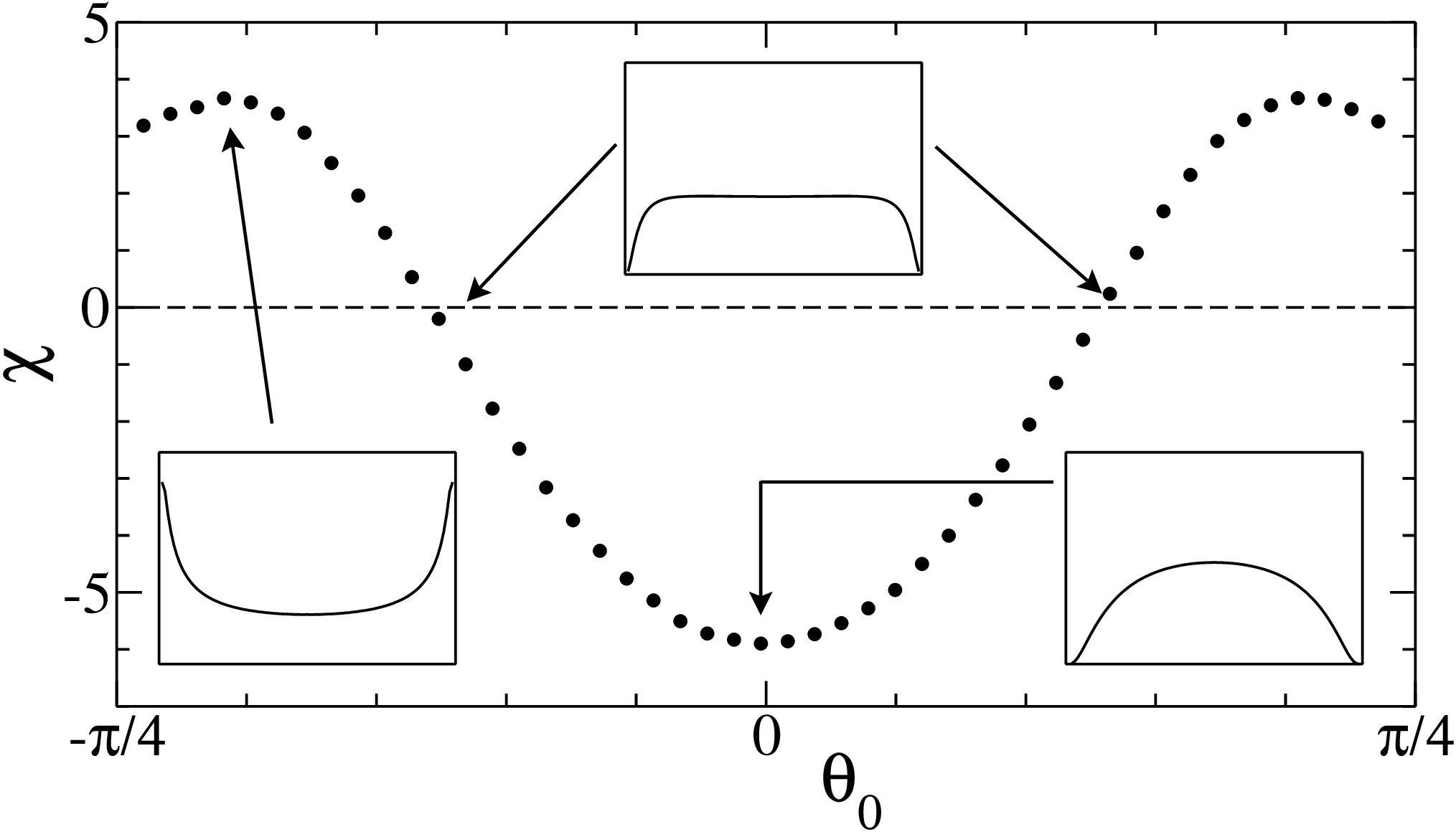}}
\caption{Dependence of the parameter $\chi$ on the starting angle
$\theta_0$ for fixed $\rho_0=0.76$ in a pie-wedge domain with opening
angle $\Theta=\pi/2$. The three insets show the shape of the uniformity
distribution $P(\omega)$ for the $\theta_0$ values indicated by the arrows.}
\label{fig:trans}
\end{figure}

To quantify the shape of the distribution of $\omega$ we perform a fit of
the numerically obtained $P(\omega)$ to a quadratic polynomial of $\omega$
in the domain $0.05<\omega<0.95$. From this fit we obtain the coefficient
$\chi$ of the quadratic term. The sign of $\chi$ thus determines the shape
of $P(\omega)$: $\chi<0$ corresponds to the unimodal, bell-shaped distribution,
$\chi>0$ signifies that the distribution is bimodal, M-shaped, and a zero value
of $\chi=0$ means that $P(\omega)$ is uniform. Such a procedure, of course, has
some ambiguities, especially when we deal with the demarcation line $\chi=0$
between regions in which $P(\omega)$ has unimodal and bimodal forms, as it is
not always clear how many digits are to be taken into account. This results
in a certain broadening of the demarcation line. However, we have checked in
several cases that this procedure produces reliable results. In
Fig.~\ref{fig:trans} we show the evolution of $\chi$ versus the starting angle
$\theta_0$ for fixed $\rho_0=0.76$ in a pie-wedge domain with the opening angle
$\Theta=\pi/2$. We observe a continuous, periodic variation of $P(\omega)$
with the starting position, changing from an M-shaped to a bell-shaped form.
The insets of Fig.~\ref{fig:trans} show the schematic distribution $P(\omega)$
for some specific values of $\theta_0$. The absolute value of $\chi$ indicates
how far the distribution $P(\omega)$ deviates from a locally uniform
distribution.

\begin{figure}
\centerline{\includegraphics*[scale=0.12]{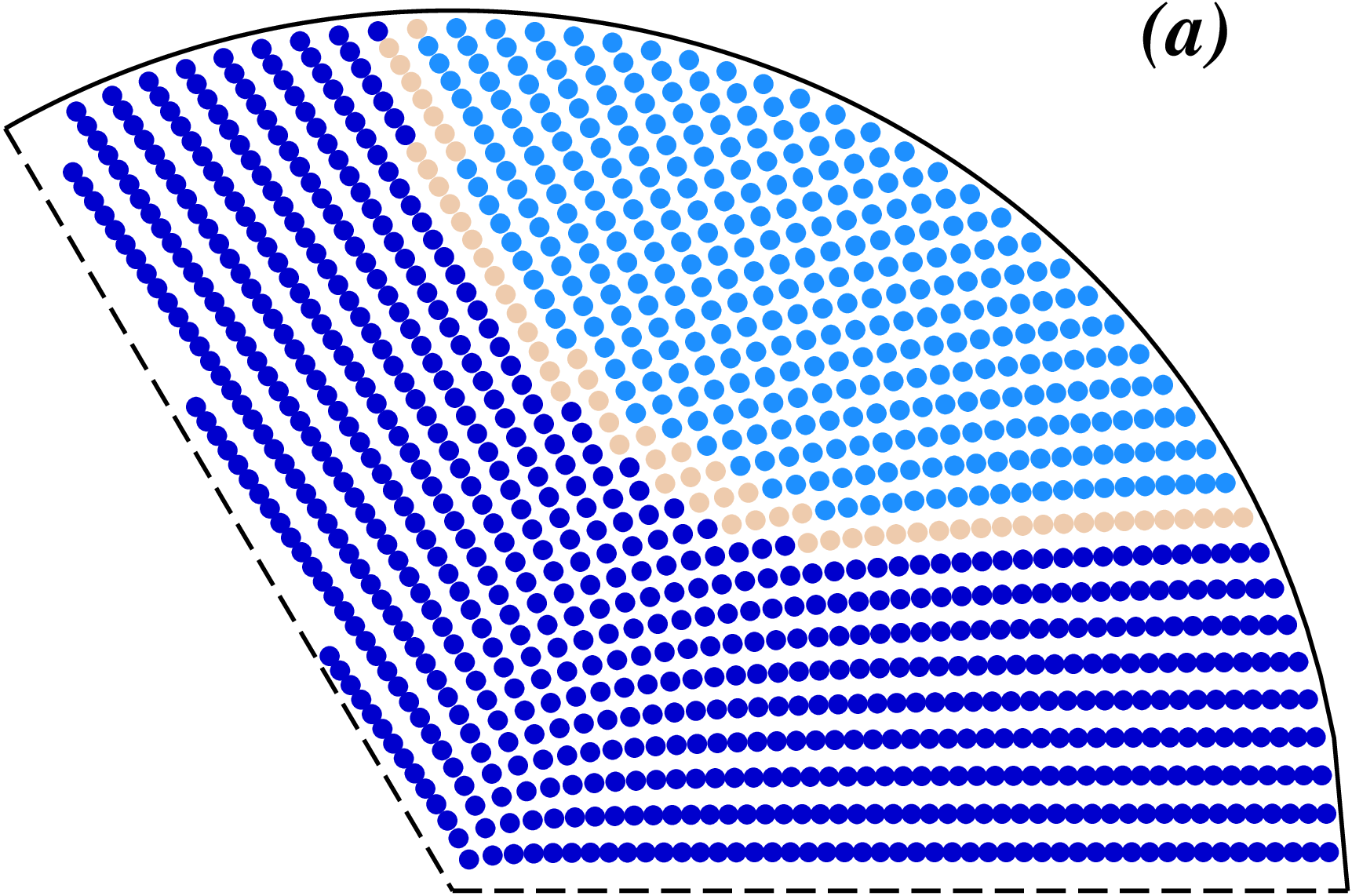}~\includegraphics*[scale=0.12]{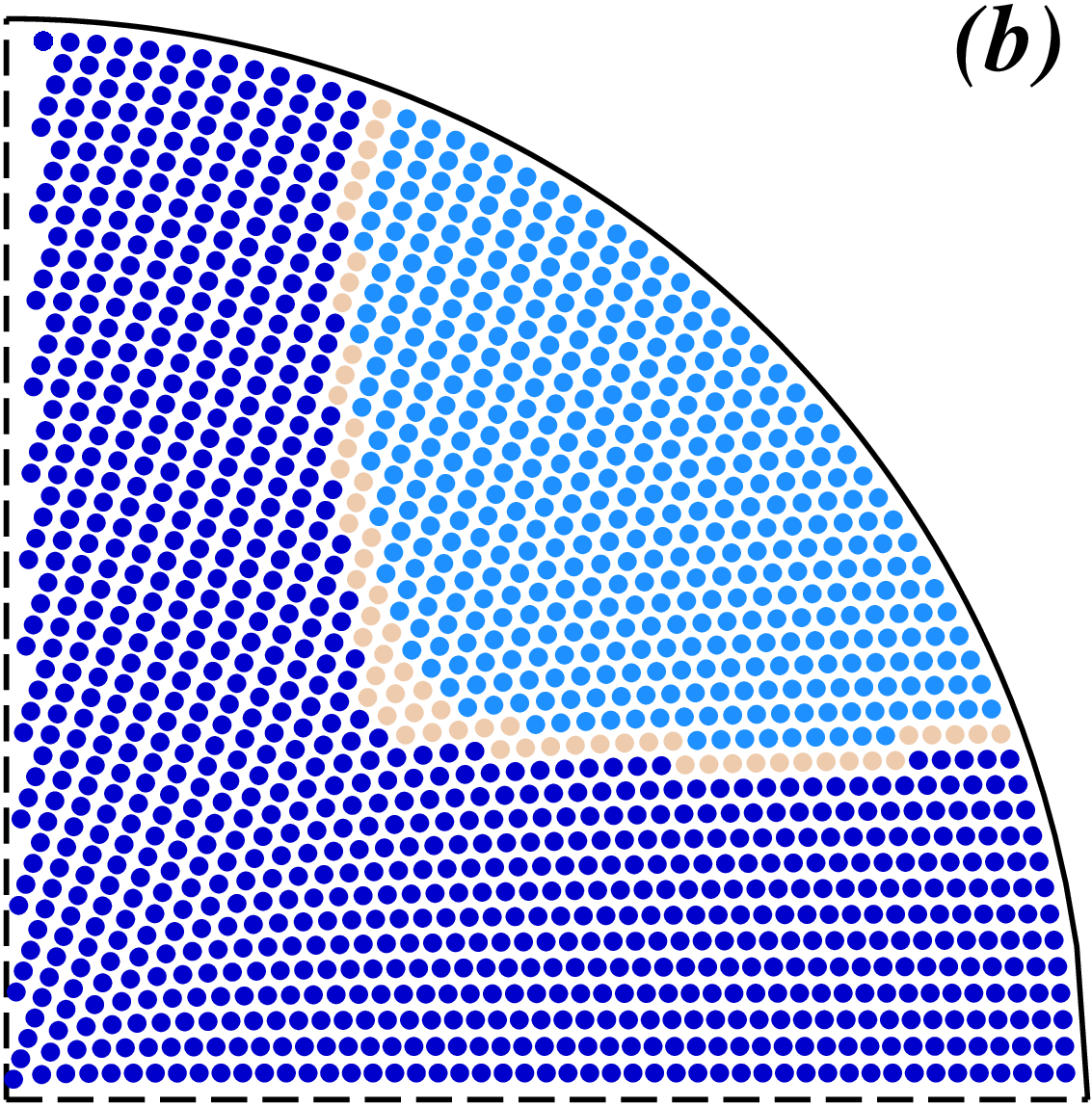}~\includegraphics*[scale=0.12]{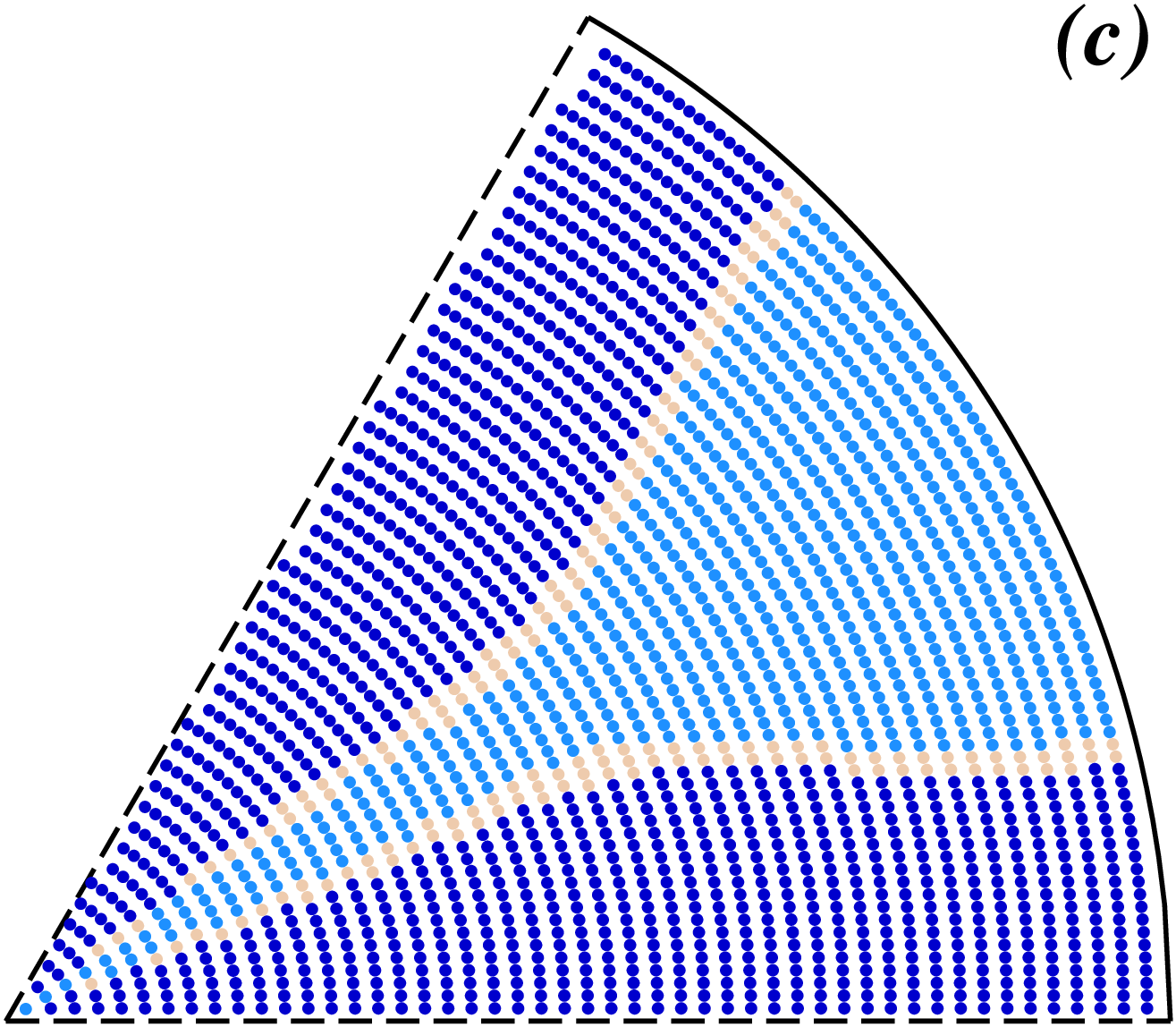}}
\caption{(Color   online)  Phase-chart for the shape of the uniformity
distribution $P(\omega)$ in three different pie-wedge domains: (a) $\Theta=
3\pi/4$, (b) $\Theta=\pi/2$, (c) $\Theta=\pi/3$. The color of the symbols
is light blue if $\chi<-\chi_\star$, dark blue if $\chi>\chi_\star$, and beige if
$|\chi|<\chi_\star$, where we chose $\chi_\star=0.25$. For each initial
location, $P(\omega)$ was computed from a sample of $N=10^5$ random
trajectories.}
\label{fig:pie-phase}
\end{figure}

Finally, we used this approach to create the phase-chart for the shape of
the uniformity distribution $P(\omega)$ with respect to the starting position
of the walker within the pie-wedge domain. In Fig.~\ref{fig:pie-phase} we
present a systematic scan of the domain for three pie-wedges with different
opening angles. One observes that in all three cases, there exists a region
in which $P(\omega)$ is bell-shaped (light blue symbols) and a region with M-shaped
$P(\omega)$ (dark blue symbols), separated by a small region with nearly uniform
distribution (beige symbols).

Therefore, as we have already remarked, the actual behavior in a finite
pie-wedge appears to be much richer than in an infinite wedge. Consider
an experiment in which one aims to find an estimate of the MFPT by tracking
the evolution of a few single particle trajectories starting at the same
position inside the light blue region. The outcome of such an experiment will be a
good estimate of the MFPT, with reliably small error. This will be the
case since in the light blue region $P(\omega)$ is bell-shaped, which means
that the probability that the two trajectories arrive at the same time
is maximal. In contrast, if two single particle trajectories start
anywhere inside the dark blue region, then it is most likely that these
trajectories will arrive to the adsorbing boundary at very different
times, yielding a poor and unreliable estimate for the MFPT. The
sample-to-sample fluctuations in this case are very important and, as
a consequence, the MFPT is not an adequate measure of the actual behavior. 
Qualitatively, the
sample-to-sample fluctuations of the MFPT increase as the trajectories
start closer to the absorbing boundaries. However, this is not always
true, as can be observed in Fig.~\ref{fig:pie-phase} (c) for the
pie-wedge with $\Theta=\pi/3$ for which the light blue region extends toward
the vertex of the wedge.

\section{Circular domain with aperture}
\label{sec:circle}

We now turn our attention to the first passage time problem of a Brownian
particle in a circular domain of unit radius and the following boundary
conditions: the segment with $|\theta|<\Theta/2$ is absorbing while the
remaining part of the outer circle is reflective. The aperture of the
circular domain is thus of angle $\Theta$. One often encounters a
three-dimensional version of this problem in cellular biochemistry, when
one is interested in the time needed for a particle (a ligand, etc.),
diffusing within a bounded domain (for instance, a microvesicle)
to reach a small escape window or a binding site, which is an aperture
in an otherwise reflecting boundary. This is the so-called Narrow Escape
Time problem, which attracted considerable attention within the last two
decades (see, e.g., Refs.\cite{ol4,NET} and references therein).

\begin{figure}
\begin{center}
\begin{tabular}{cc}
\includegraphics*[width=0.32\linewidth]{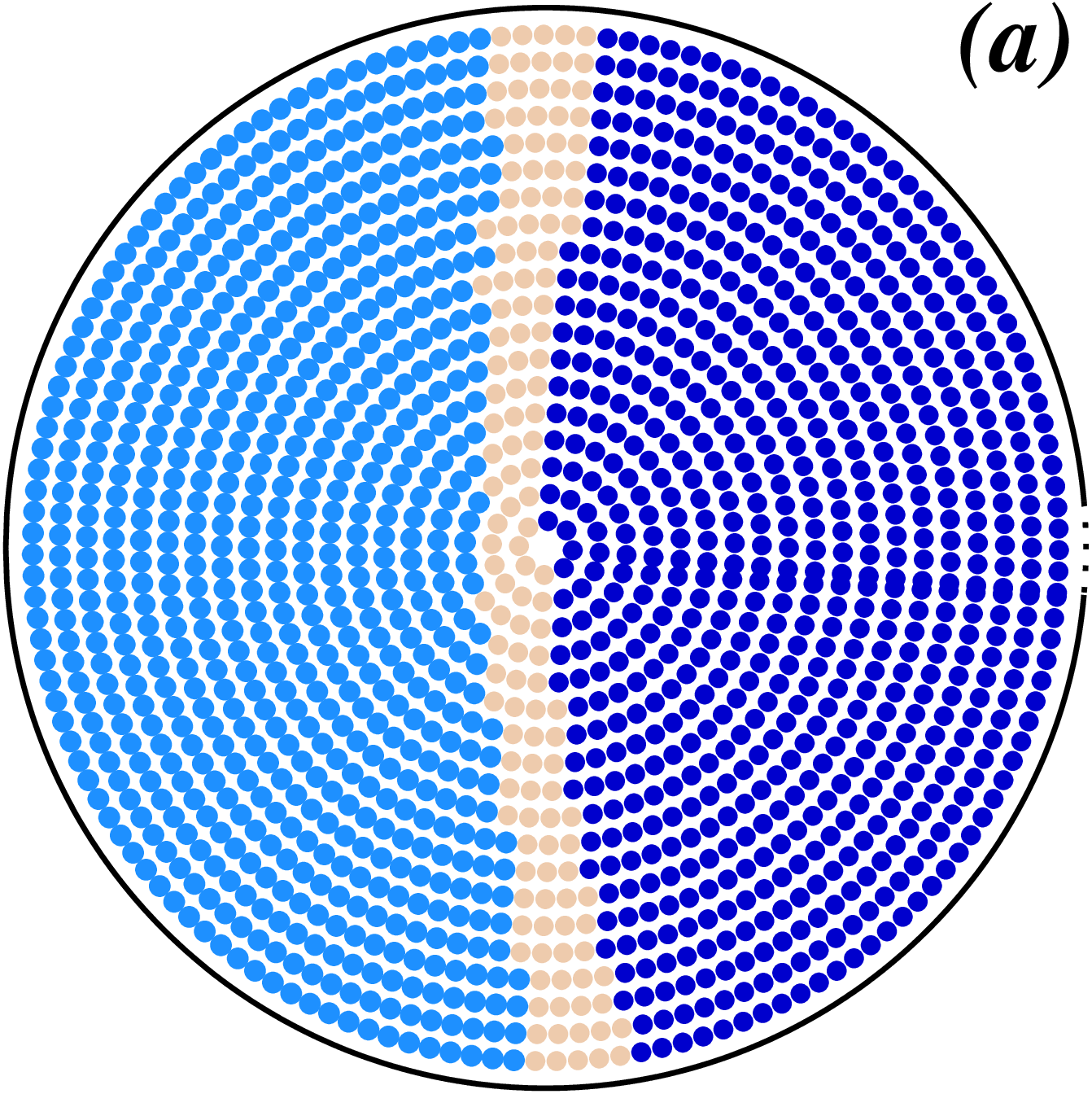} & \includegraphics*[width=0.32\linewidth]{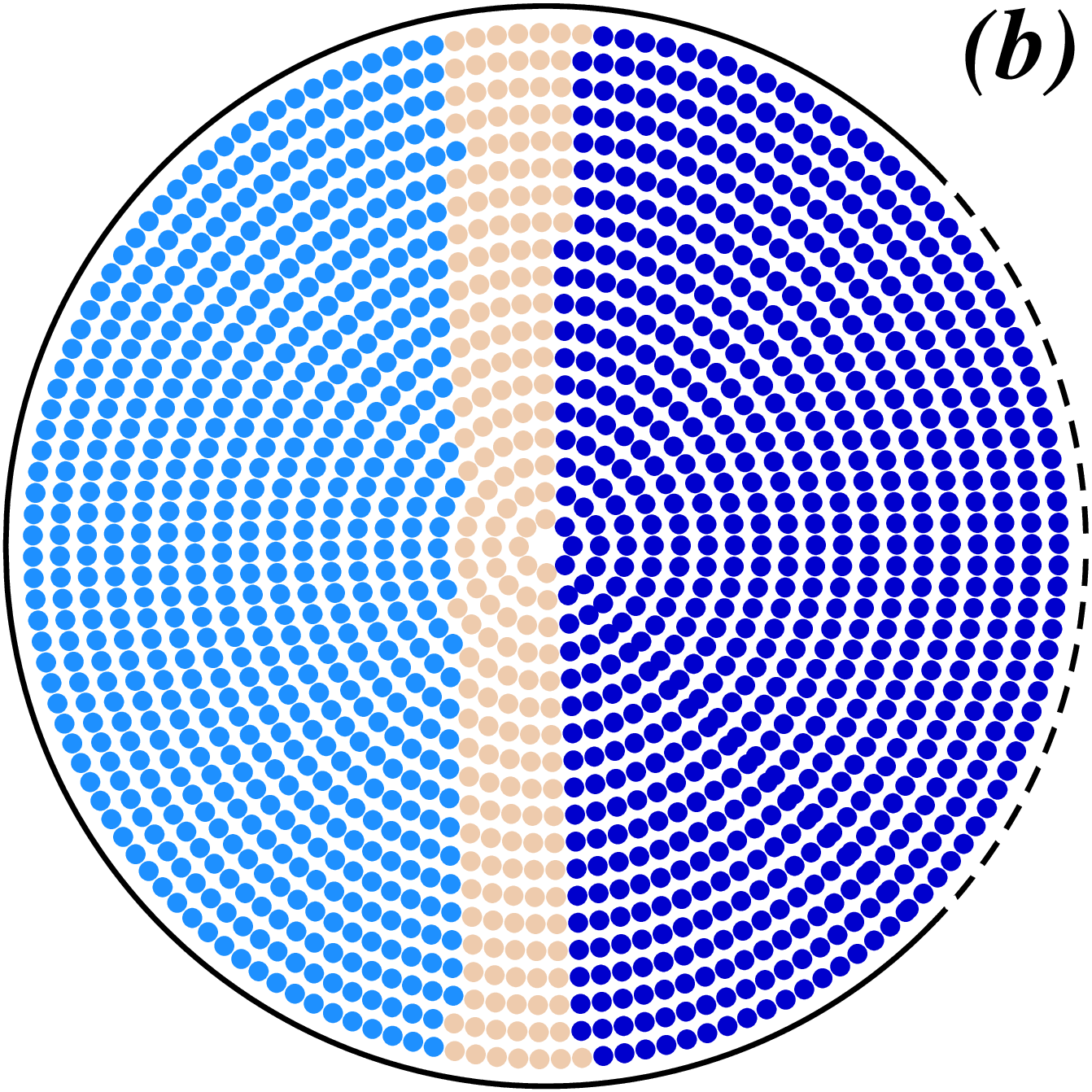} \\
\includegraphics*[width=0.32\linewidth]{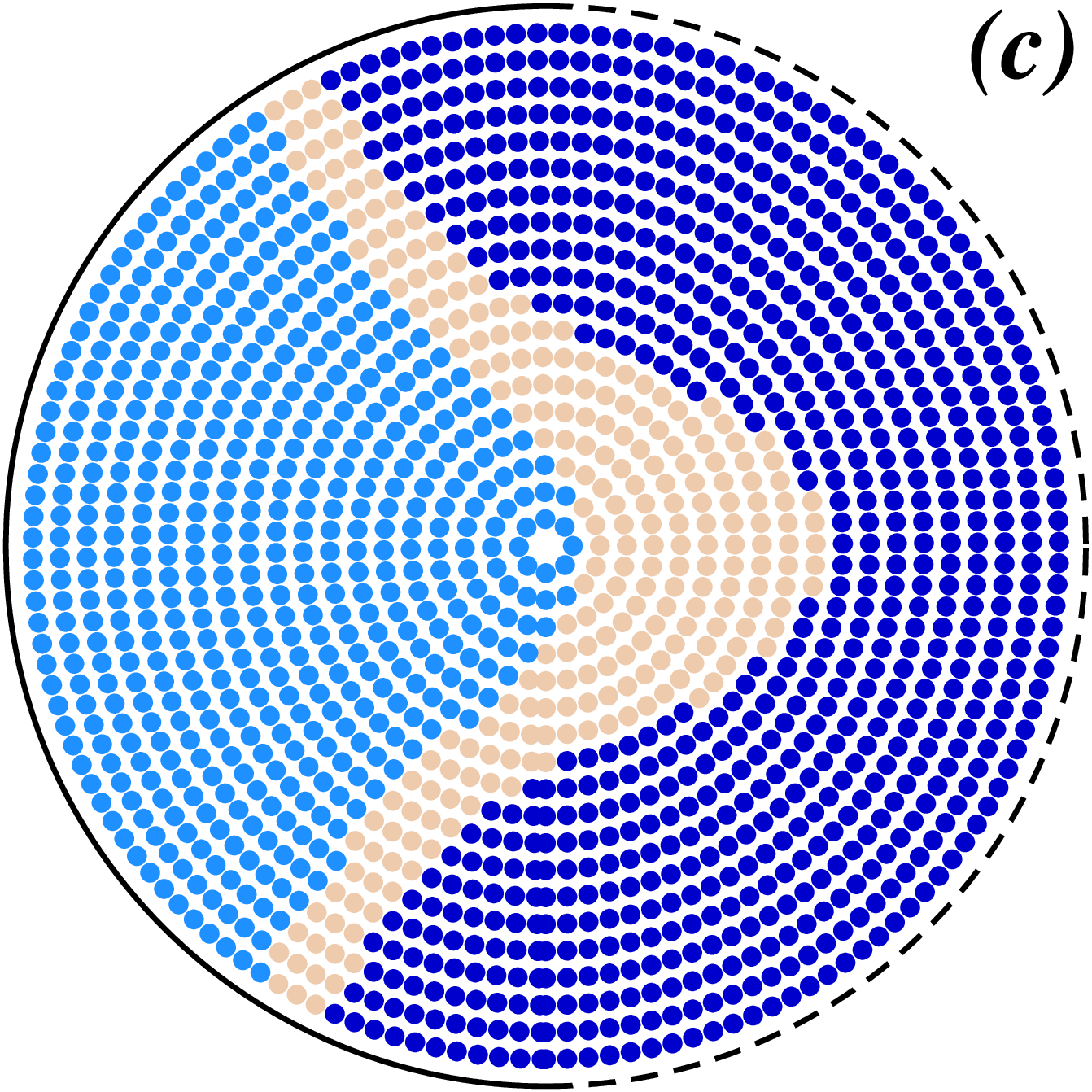} & \includegraphics*[width=0.32\linewidth]{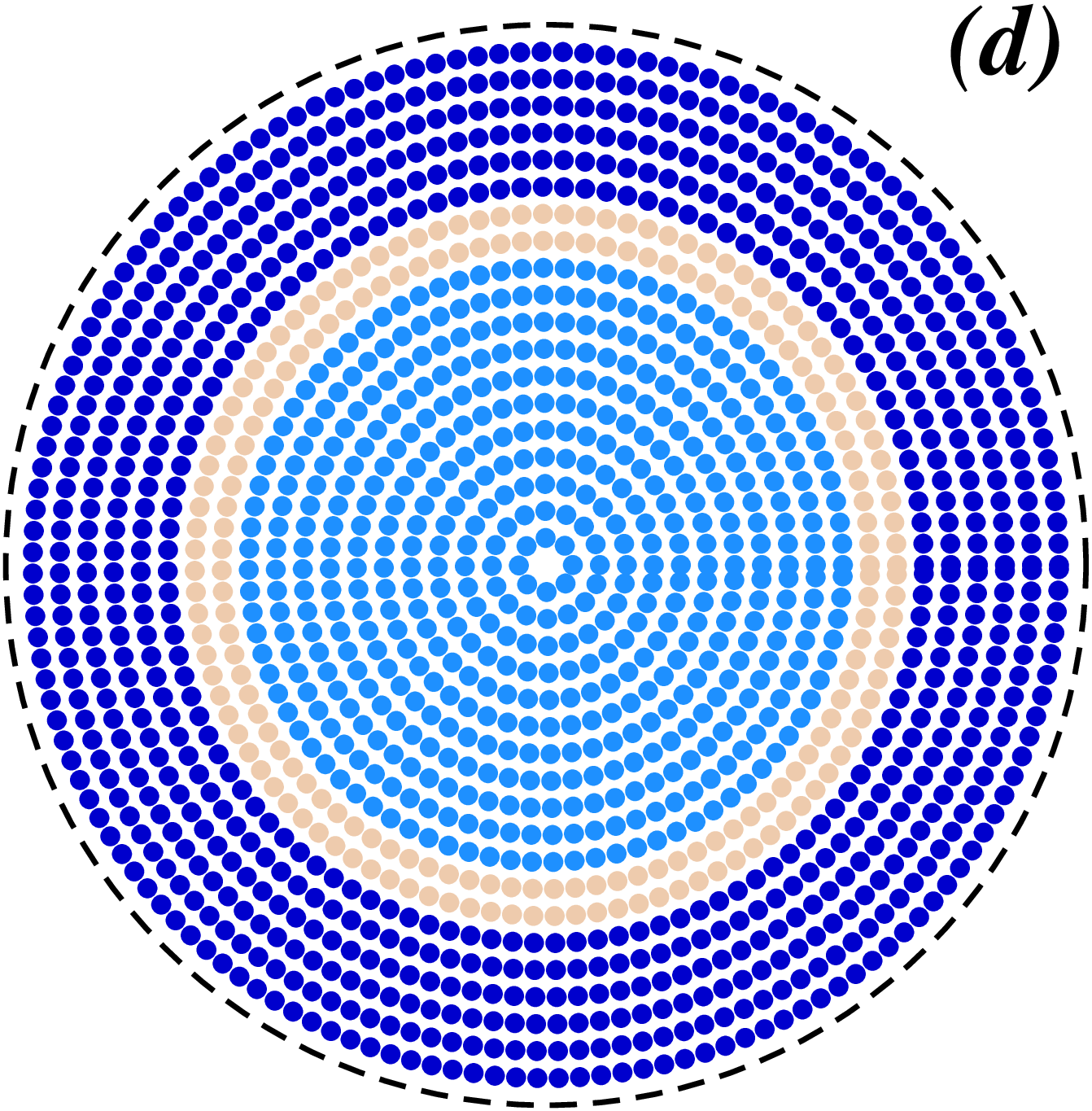} \\
\end{tabular}
\includegraphics*[width=0.7\linewidth]{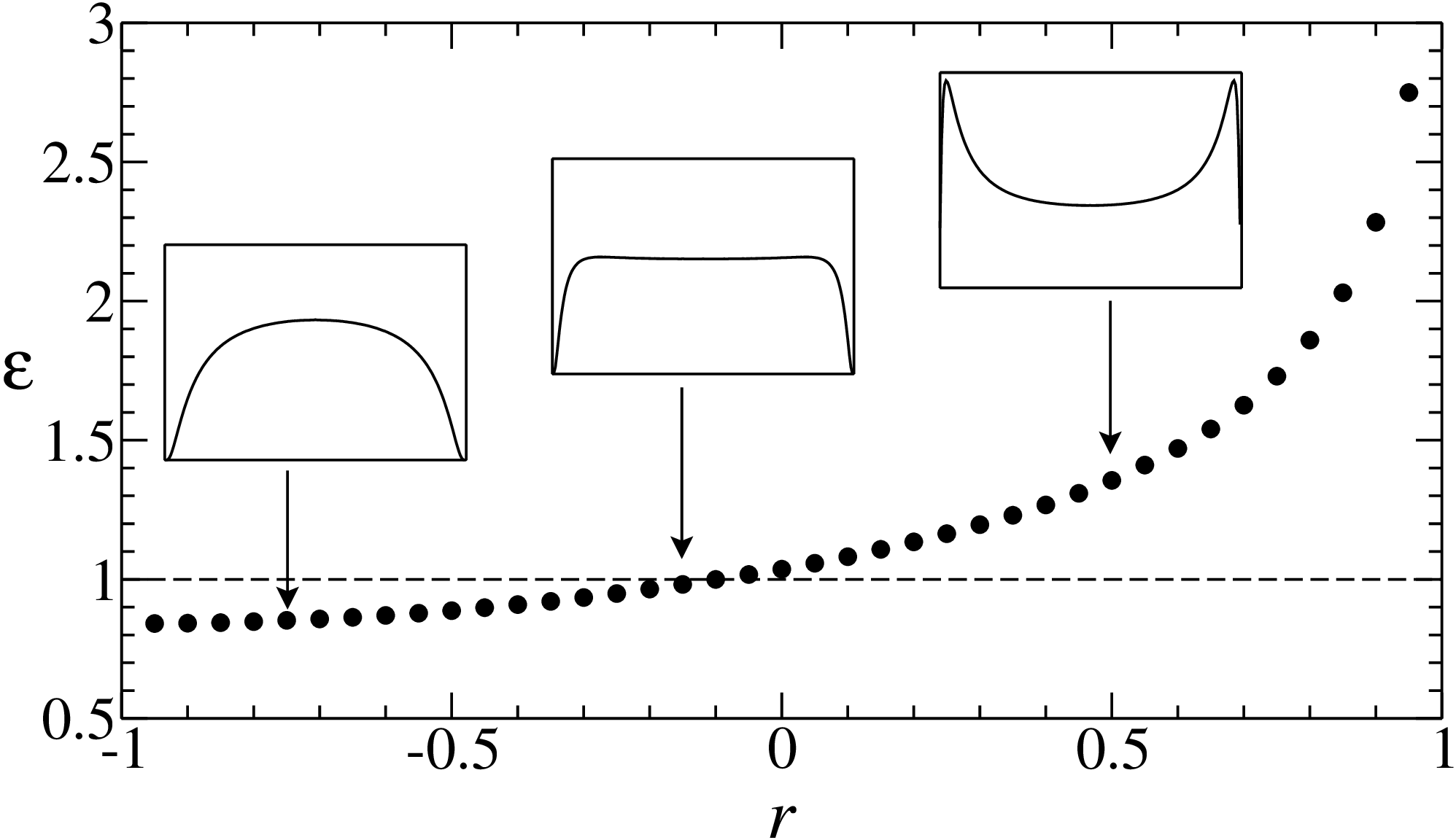}
\end{center}
\caption{(Color online) Phase-chart for the shape of the uniformity
distribution $P(\omega)$ for a Brownian walker in the unit circle with reflective
BCs (solid lines) and an aperture of size $\Theta$ (absorbing BC - dashed lines). (a) $\Theta=\pi/18$,
(b) $\Theta=\pi/2$, (c) $\Theta=\pi$, and (d) $\Theta=2\pi$. Starting locations are colored light blue
if $\chi<-\chi_\star$, dark blue if $\chi>\chi_\star$ and beige if $|\chi|<\chi_
\star$, where $\chi_\star=1$. In the lower panel, the relative error of the
FPT $\varepsilon$ is shown as function of the initial radial position $r$ 
along the horizontal diameter of the unit circle with $\Theta=\pi/2$, panel
(b): $r=0$ is associated to the center of the circle and the point $r=1$
lies over the absorbing boundary.}
\label{fig:circle}
\end{figure}

We analyze the shape of the uniformity distribution $P(\omega)$ as a function
of the starting point of a Brownian walker. As in the previous section, we
generate $N=10^5$ random walks commencing from the same starting position
$(\rho_0,\theta_0)$ inside the unit circle and determine the set $\{\tau_i\}$
of first passage times to the location of the aperture. From these data we
obtain $P(\omega)$ and compute the parameter $\chi$. In Fig.~\ref{fig:circle}
we show  the phase-chart for the shape of $P(\omega)$ for four different sizes
of the aperture: $\Theta=\pi/18$, $\pi/2$, $\pi$, and $2\pi$. Each symbol in
the charts is light blue, beige, or dark blue, depending on whether the corresponding
starting position leads to a bell-shaped, uniform, or M-shaped distribution
$P(\omega)$, respectively. Note that the case shown in Fig.~\ref{fig:circle}
(d) reduces to a one-dimensional problem (see, e.g., Ref.~\cite{carlos}).

Similarly to our findings for the pie-wedge domain, we observe that
the MFPT is not always a representative measure for the
two-dimensional Narrow Escape Time problem.  Interestingly, the sub-domain
in which the MFPT is the least probable outcome (dark blue coding) is
practically the same for small holes of $\Theta\le\pi/2$.  It is
worthwhile noting that while in this region $P(\omega)$ is always
bimodal, the height of its maxima increases (and its value
$P(\omega=1/2)$ representative of the MFPT decreases) depending on the
distance from the opening.

In addition, from the set of first passage times $\{\tau_i\}$ we
directly computed the MFPT and the variance $\mathrm{var}(\tau)$. Both
statistical indicators grow with the distance from the absorbing
boundary. A more sensitive measure is the relative error
$\varepsilon$, defined as the ratio
\begin{equation}
\label{eq:eps}
\varepsilon=\frac{\sqrt{\mathrm{var}(\tau)}}{\langle\tau\rangle}.
\end{equation}
In the lower panel of Fig.~\ref{fig:circle} we show the dependence of
the relative error on the starting position $\mathbf{r}_0$ for trajectories
initiating along the symmetry axis of the domain, namely with respect to
$r_0$ for fixed $\theta_0=0$. In agreement with the qualitative results of
the phase-chart, $\varepsilon<1$ only when $P(\omega)$ is bell-shaped, and
the MFPT is the most probable outcome of a single-particle trajectory, namely,
for trajectories starting far enough from the absorbing boundary. Clearly, the
closer the starting position is to the absorbing boundary the larger the
relative error becomes. Very near the absorbing boundary the standard
deviation of the first passage time becomes much larger than its mean.
We note that this result is generic irrespectively of the aperture size.

However $\varepsilon$ is just a number and it is not clear how to interpret it.
For instance, $\varepsilon = 1$ or $\varepsilon = 2$, are these values too small or large
enough to allow us to say that trajectory-to-trajectory fluctuations are significant?
On the other hand, the distribution of the simultaneity index which we discuss here
gives a lucid answer on this question as manifested by the change of modality of $P(\omega )$.

\section{Triangular domain with absorbing boundaries}
\label{sec:triangle}

\begin{figure}
\centerline{\includegraphics*[width=0.3\linewidth]{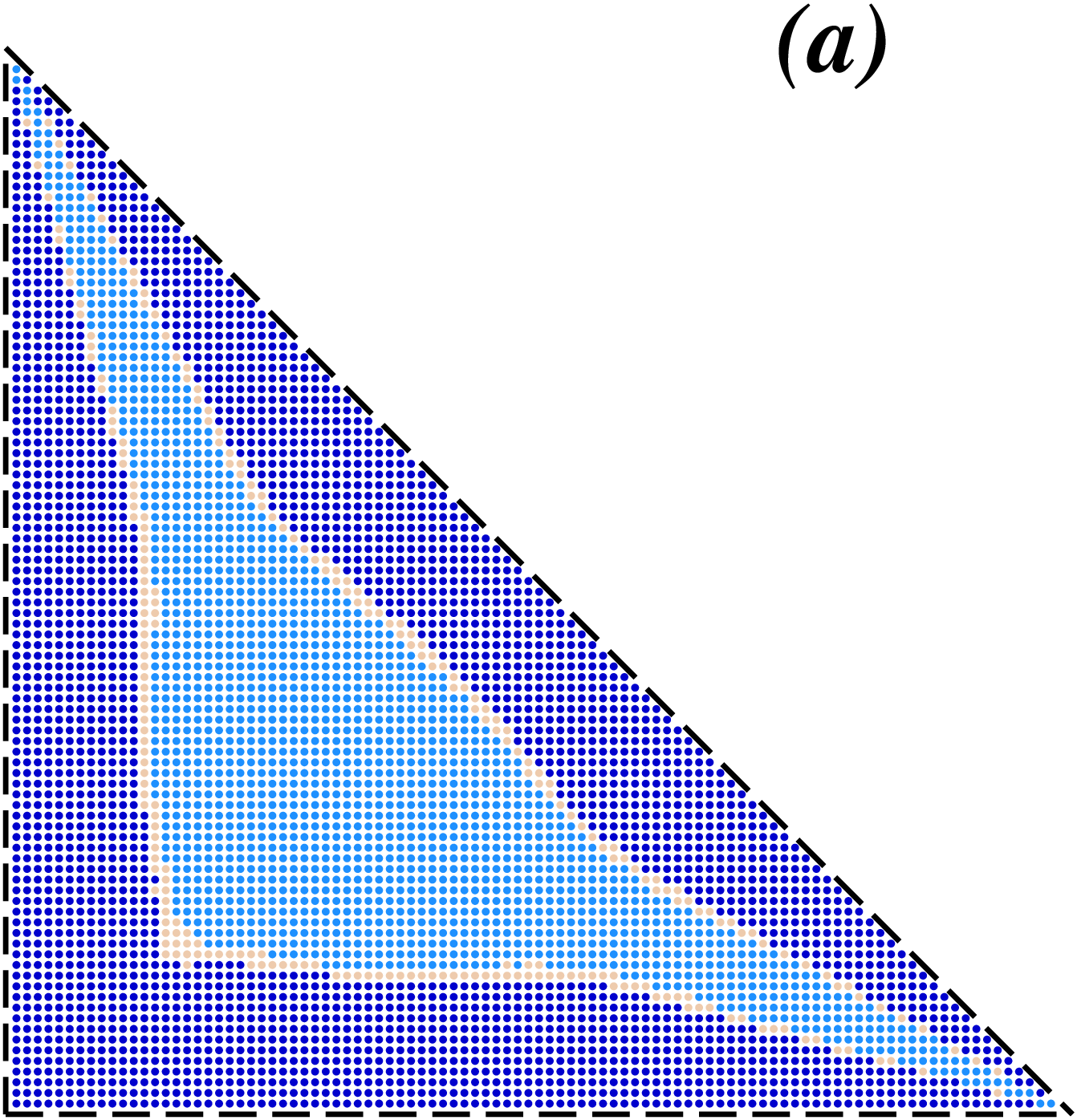}~\includegraphics*[width=0.3\linewidth]{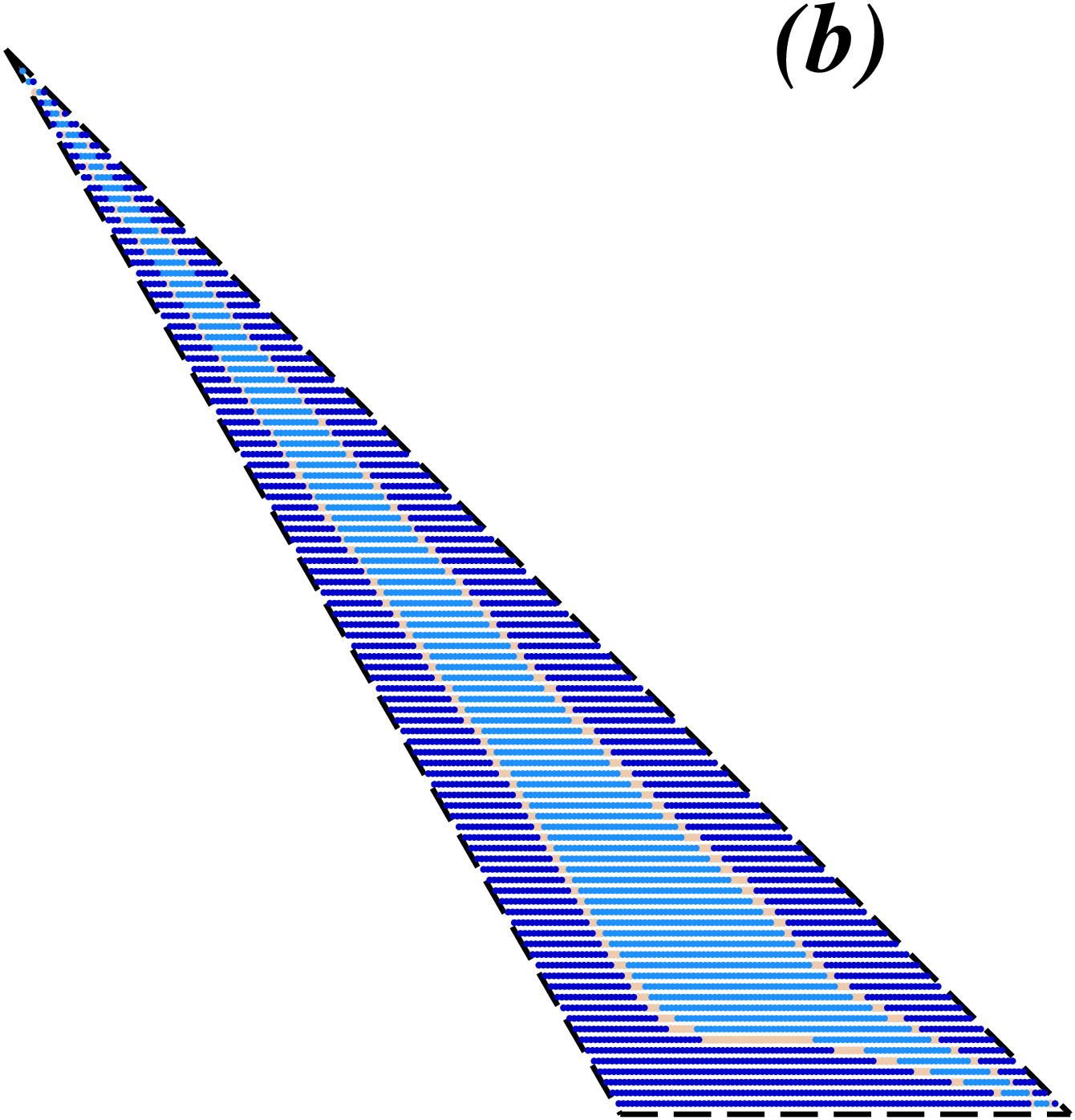}~\includegraphics*[width=0.396\linewidth]{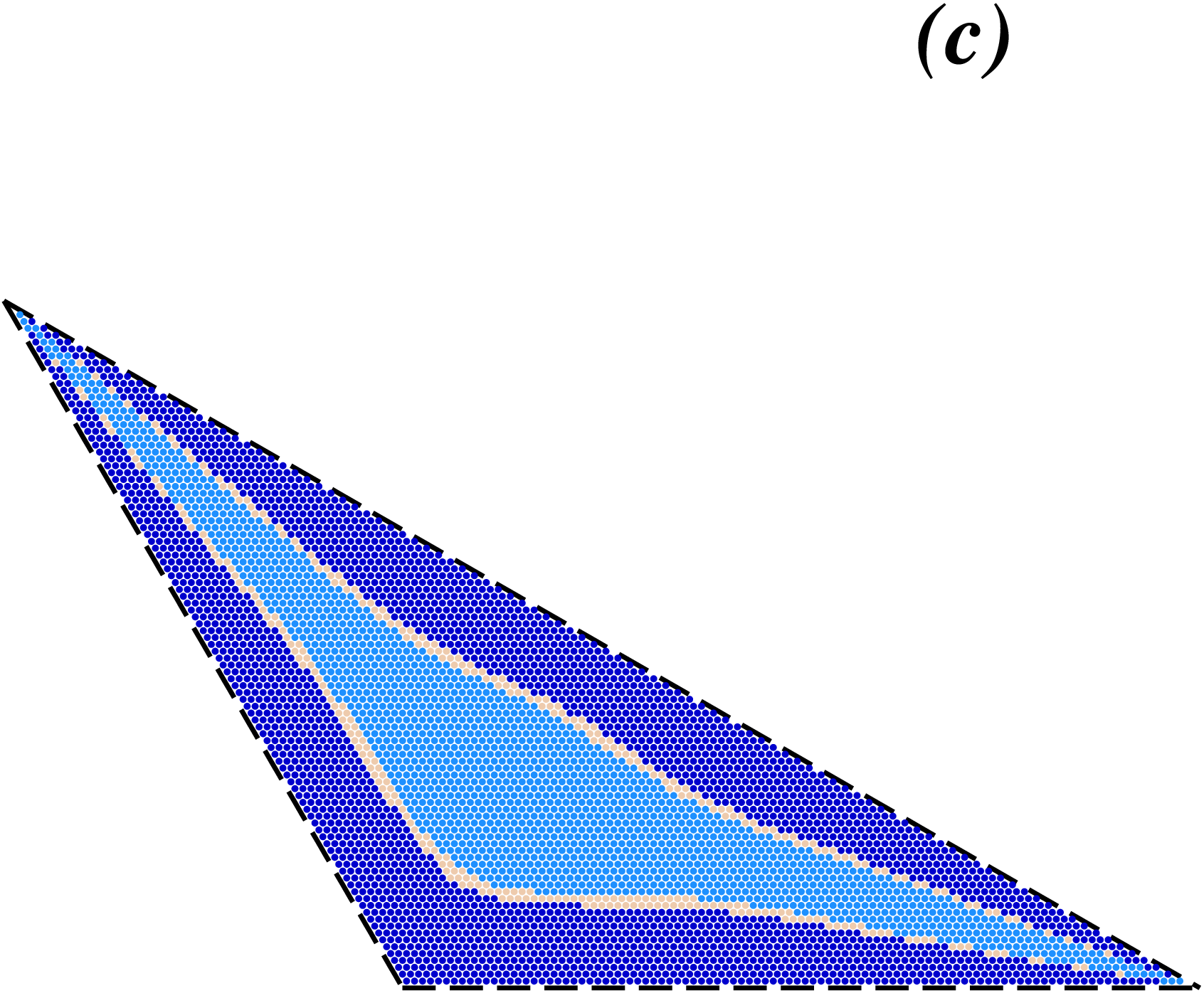}}
\caption{(Color online) Phase-chart of the shape of the uniformity
distribution $P(\omega)$ for symmetric triangles with central angle
$\pi/2$ (a) and $2\pi/3$ (c), and for the asymmetric triangle with
angles $2\pi/3,\pi/4,\pi/12$ (b). The color of the symbols is light blue
if $\chi<-\chi_\star$, dark blue if $\chi>\chi_\star$, and beige if
$|\chi|<\chi_\star$, with $\chi_\star=0.25$.}
\label{fig:triangle}
\end{figure}

Finally, as a complementary example we consider a domain whose boundaries
are completely absorbing. We consider the triangular domains shown in
Fig.~\ref{fig:triangle}: two symmetric triangles with central angle $\pi/2$,
panel (a), and $2\pi/3$, in panel (c), and an asymmetric triangle with angles
$2\pi/3,\pi/4,\pi/12$ shown in panel (b).

In Fig.~\ref{fig:triangle} we show the phase-chart of the shape of
$P(\omega)$, the  results obtained are qualitatively the same as for the
previous two examples with mixed boundary conditions. $P(\omega)$ is
bell-shaped and the MFPT the most probable outcome only when the trajectory
starts far enough from the boundary. Also, similarly to  what we observed
for the pie-wedge domain, the domain in which $P(\omega)$ is unimodal extends
toward the absorbing boundary if the vertex angle is less than $\pi/2$.

\section{Conclusions}
\label{sec:concl}

We explored the problem of first passage of a Brownian particle to the
absorbing boundary of finite, two-dimensional domains. From our study
of the characteristic shapes of the associated distribution of the
uniformity index $\omega$ we demonstrated that the MFPT represents the
most probable outcome (and thus is quite meaningful) 
only if the trajectories start in a certain subregion
of the total domain. For starting points in the complementary region the
MFPT becomes the least probable outcome, indicating very large sample-to-sample
fluctuations. These observations are generically important for single
trajectory analysis of first passage time processes.

We showed that the associated separation into bell-shaped and M-shaped
forms of the uniformity distribution $P(\omega)$ is a robust property
of Brownian motion by studying the problem in different symmetric and
asymmetric domains with mixed or fully absorbing boundaries. We found
that in general, sample-to-sample fluctuations of the first passage
time increase when the trajectories start close to the target
boundary, leading to the unexpected conclusion that in such situations
the MFPT yields insufficient information, particularly, if the
absorption time is extracted from the outcome of very few
single-particle trajectories.

Next, it is worthwhile mentioning that in many interesting
situations the starting position of the trajectories are randomly
distributed inside the finite domain. From such analysis the so-called global
MFPT is usually derived, see, e.g., Ref.~\cite{ol4}. Here we found that
averaging the associated uniformity distribution $P(\omega)$ over the domain,
\begin{equation}
P_{\mathrm{av}}(\omega)=\int_{\mathcal{S}} P_{\mathbf{r}_0}(\omega)
d\mathbf{r}_0,
\end{equation}
attains a uniform shape, except near $\omega=0$ and $\omega=1$. This
appears to be a general property of $P_{\mathrm{av}}(\omega)$ associated
with the probability conservation, and leads to the unexpected conclusion
that the global MFPT has little meaning in such situations.

As a final remark, we emphasize that the approach outlined here
is not limited to first passage phenomena only, but can be quite generally
applied to probe the significance of sample-to-sample
fluctuations of arbitrary random variables having distributions for which
\textit{all\/} moments exist. Such distributions, as shown in our work,
may appear $\omega$-broad, in the sense that the corresponding uniformity
distribution $P(\omega)$ is bimodal, or, alternatively, $\omega$-narrow
with unimodal $P(\omega)$.
We recall that the variable $\omega$ has a very lucid physical
meaning and its distribution can be determined if the parental distribution
of the random variable is known. Indeed such sort of heterogeneity analysis has very recently started to be used to quantify sample-to-sample fluctuations in mathematical finances~\cite{samor2,hol}, chaotic systems~\cite{schehr}, analysis of distributions of the diffusion coefficient of proteins diffusing along DNAs~\cite{boyer} and FPT phenomena~\cite{carlos}.

\begin{center}
\textbf{ACKNOWLEDGMENTS}
\end{center}

The authors wish to thank Olivier B\'enichou and Satya Majumdar
for helpful discussion.
The authors acknowledge financial support from the European Science
Foundation and the  hospitality of NORDITA, Stockholm, where part of
this work was performed during the Non-Equilibrium Statistical
Mechanics program. The research of TGM, RM and GO is partially
supported by a Marie Curie International Research Staff Exchange
Scheme Fellowship PIRSES-GA-2010-269139 within the 7th European Community
Framework Programme. RM acknowledges funding from the Academy of
Finland within the FiDiPro programme. The authors are
partially supported by the ESF Research Network "Exploring the Physics
of Small Devices".


\end{document}